\documentclass[11pt,a4paper]{article}

\pdfoutput=1

\usepackage{jcappub}
\usepackage{graphicx}
\usepackage{float}
\usepackage{slashed}
\usepackage[font=small, labelfont=bf]{caption}
\usepackage[labelformat=simple]{subcaption}

\usepackage{amsmath, amsthm, amssymb, amsfonts}
\usepackage{multirow}
\usepackage[numbers,sort&compress]{natbib}
\usepackage{hyperref}

\usepackage{amsmath}
\usepackage{amssymb}
\usepackage{graphicx}
\usepackage{color}
\usepackage{hyperref}

\def\be{\begin{equation}}
\def\ee{\end{equation}}
\def\bea{\begin{eqnarray}}
\def\eea{\end{eqnarray}}

\begin{document}

\title{A geometrical instability for ultra-light fields during inflation?}

\author[a,b,c]{Michele Cicoli,}
\author[a,b]{Veronica Guidetti,}
\author[a,b]{Francisco G. Pedro,}
\author[b]{Gian Paolo Vacca}

\affiliation[a]{\small Dipartimento di Fisica e Astronomia, Universit\`a di Bologna, via Irnerio 46, 40126 Bologna, Italy}
\affiliation[b]{\small INFN, Sezione di Bologna, viale Berti Pichat 6/2, 40127 Bologna, Italy}
\affiliation[c]{\small  Abdus Salam ICTP, Strada Costiera 11, Trieste 34151, Italy}

\emailAdd{mcicoli@ictp.it}
\emailAdd{veronica.guidetti2@unibo.it}
\emailAdd{francisco.pedro@bo.infn.it}
\emailAdd{gianpaolo.vacca@bo.infn.it}

\abstract{
We study the behaviour of isocurvature perturbations in non-linear sigma models which naturally emerge in supergravity and string inflationary scenarios. We focus on the case of negatively curved field manifolds which can potentially lead to a geometrical destabilisation of isocurvature modes. We find however that heavy fields are stable when their effective mass is computed on the attractor background solution. On the other hand, we show that ultra-light fields can potentially suffer from a geometrical instability when the background trajectory is a geodesic in field space. In this case a full understanding of the system is likely to require the use of non-perturbative methods.
}

\maketitle

\section{Introduction}

Many inflationary scenarios beyond the SM feature non-linear sigma models characterised by multiple scalar fields and a curved field manifold. In particular, these arise naturally within the framework of supergravity, string compactifications and models with non-minimal coupling. In a multi-field set-up, there are several spectator fields which can be either \textit{heavy}, i.e. $m_{\rm h}\gg H$, or \textit{light}, i.e. $m_{\rm l}\ll H$. When the field manifold is negatively curved, the effective mass of these isocurvature modes receives negative contributions which can potentially induce a geometrical instability by making them tachyonic \cite{Renaux-Petel:2015mga}. In this paper we show however in Sec. \ref{sec:inst} that heavy modes are stable when the system evolves along the attractor background trajectory,  in agreement with previous results found in models with non-minimal coupling  \cite{ Kaiser:2012ak,Schutz:2013fua}. In Sec. \ref{sec:lightscalars} we present our main result which is the new observation that a potential geometrical instability arises instead generically for ultra-light fields, i.e. $m_l \to 0$, when the background trajectory is geodesic. In this case a full understanding of the inflationary dynamics requires going beyond perturbation theory.
\section{Geometrical stability for heavy fields}
\label{sec:inst}

The Lagrangian of a generic non-linear sigma model is:
\be
\mathcal{L}/\sqrt{|g|}=\frac12 G_{ij}(\phi_i)\partial_\mu \phi_i \partial^\mu \phi_j-V(\phi_i)\,,
\label{eq:L}
\ee
where $G_{ij}(\phi_i)$ denotes the field space metric. In such multi-field models the background trajectory defines a projection for the gauge invariant perturbations into a tangent component, the curvature perturbations, and an orthogonal component, the isocurvature perturbations. The inflationary dynamics of these models has been intensively studied over the last two decades and it has been shown to be significantly richer than that of single-field models while still being compatible with observational constraints. In this paper we are interested in the super-horizon behaviour of the isocurvature perturbations which is determined by their effective mass (setting $M_p=1/\sqrt{8\pi G}=1$) \cite{Sasaki:1995aw, DiMarco:2002eb,Achucarro:2010da}:
\be
m_{\rm eff}^2\equiv N^i N^j \left(V_{ij}-\Gamma_{ij}^k V_k\right)+\left(\epsilon \mathbb{R}+3\eta_\perp^2\right) H^2\,,
\label{eq:meff2}
\ee
where $T^i$ and $N^i$ are the unit vectors along the tangent and normal directions, $\mathbb{R}$ denotes the field space Ricci scalar and $\eta_\perp=N^i V_i/(H |\dot\phi|)$ measures the turning rate of the trajectory of the homogeneous background fields for which 
$
|\dot\phi|=\sqrt{G_{ij} \dot \phi_i \dot \phi_j}$.

It has been observed that $m_{\rm eff}^2$ can potentially become negative if $\mathbb{R}<0$ and $\eta_\perp=0$ \cite{Gong:2011uw, Renaux-Petel:2015mga}. Given that for heavy fields $N^i N^j V_{ij}\gg H^2$ and during inflation $\epsilon\ll 1$, this can happen only if $|R|\gg 1$. Ref. \cite{Renaux-Petel:2015mga}
 considered a simple model where a heavy field is coupled to the inflaton kinetic terms via a higher-order operator suppressed by $M$. In this case $|R|\simeq 4/M^2$, and so $M\ll 1$ can generate a large negative contribution to $m_{\rm eff}^2$. This could trigger an instability characterised by a super-horizon growth of the isocurvature perturbations which signals a breakdown of perturbation theory and a potential premature end of inflation \cite{Renaux-Petel:2015mga}. Let us point out that negatively curved field manifolds arise naturally both in supergravity and in multifield models with non-minimal couplings \cite{Kaiser:2010ps}. In fact, the simple K\"ahler potential $K=-3\ln(T+\bar{T})$ for the complex volume modulus $T$ gives $R=-8/3$. However the reference scale of supergravity is the Planck mass, and so $M\simeq 1$. This implies that during inflation generically $\epsilon |\mathbb{R}| H^2\ll H^2$, resulting in an absence of any geometric instability.

In what follows we shall however show that, even if $|R|\gg 1$, $m_{\rm eff}^2$ is negative only if it is computed on a repulsive trajectory, while the isocurvature modes are stable if the system evolves along the attractor trajectory. Thus the physical interpretation of $m_{\rm eff}^2<0$ is not that quantum fluctuations grow beyond the regime of validity of perturbation theory but that the classical field trajectory is unstable under perturbations of the initial conditions.

To analyse the model we focus on the case where:
\be
G_{ij}= \left(
  \begin{array}{cc}
   1 & 0  \\
0 & f^2(\phi_1) \\
  \end{array} \right)\quad \text{and} \quad V = V(\phi_1)+V(\phi_2) \,.
	\label{model}
\ee
Notice that we made this choice following \cite{Renaux-Petel:2015mga} since it allows to have simple analytic formulae but a geometrical instability can arise also for more generic cases with non-diagonal metric and non-sum-separable potential. It can be shown that the corresponding curvature scalar can be negative since it takes the form:
\be
\mathbb{R}=-2 \,\frac{f_{11}}{f}\,.
\label{eq:R}
\ee
The equations of motion from (\ref{eq:L}) and (\ref{model}) for $\phi_i=\phi_i(t)$ in an expanding Universe with $\sqrt{|g|}=a^3$ read:
\be
\dot\pi_1 = a^3 \left(f f_1\dot\phi_2^2 -V_1\right)
\qquad 
\dot\pi_2 = - a^3\, V_2\,,
\label{eq1}
\ee
where the conjugate momenta are:
\be
\pi_1 = a^3 \dot\phi_1\qquad\qquad \pi_2 = a^3 f^2 \dot\phi_2\,.
\label{eq2}
\ee
The background dynamics of the system is determined by (\ref{eq1}), (\ref{eq2}) and the Friedmann equation:
\be
H^2=\frac13\left(\frac12 G_{ij}\dot{\phi}^i\dot{\phi}^j+V\right).
\ee

\subsection{Canonical heavy field}
\label{sec:canHeavy}

We first consider the case where the heavy field has canonical kinetic terms, and so identify $\phi_1$ with the heavy scalar and $\phi_2$ with the inflaton. 
We see from (\ref{eq1}) that the equation for $\phi_1$ admits a slow-roll solution with \cite{Garcia-Saenz:2018ifx}:
\be
f f_1 \dot\phi_2^2 \simeq V_1\,,
\label{eq:attractorsolu}
\ee
which implies that $\pi_1$ is an approximately conserved quantity. Given that during inflation $a\propto e^{Ht}$, (\ref{eq2}) then gives $\dot\phi_1\to 0$. In this solution the heavy field does not sit at the minimum of its potential but it is displaced from it by the inflaton's kinetic energy. Hence the motion is non-geodesic since:
\be
\eta_\perp=\frac{V_1}{H f\dot\phi_2}=-\frac{f_1 \dot\phi_2}{H} \,,
\label{etaperp}
\ee
leading to an isocurvature mass:
\be
m_{\rm eff}^2= V_{11}+\left(3\eta_\perp^2-2 \epsilon\,\frac{f_{11}}{f}\right) H^2 \,.
\label{eq:meff2Attractor}
\ee
Using \eqref{eq:attractorsolu}, we can further simplify $m_{\rm eff}^2$ and show that on this generic solution it is strictly positive: 
\be
m_{\rm eff}^2=8 \epsilon \left(\frac{f_1}{f}\right)^2 H^2\,,
\ee
even if the Ricci scalar is negative. This signals that the background trajectory is stable, regardless of the functional form of the kinetic coupling $f^2(\phi_1)$. 

Notice however that, in particular cases where $V_1=0$ and $f f_1=0$ have a common root, (\ref{eq:attractorsolu}) could also be exactly satisfied with the heavy field sitting at the bottom of its potential. In this case $\eta_\perp=0$, and so the effective mass (\ref{eq:meff2Attractor}) reduces to:
\be
m_{\rm eff}^2=V_{11}-2\epsilon\, \frac{f_{11}}{f} H^2\,.
\label{eq:meff2Trivial}
\ee
If $0<f\ll f_{11}$ (or $f_{11}\ll f<0$), this effective mass can become negative in regimes where the field space curvature contribution dominates \cite{Gong:2011uw,Renaux-Petel:2015mga}.  

However we will now show that this unusual behaviour is merely a consequence of doing cosmological perturbation theory on a repulsive background trajectory since the trivial solution is unstable under perturbations of the initial conditions: 
\be
\phi_1=\bar{\phi}_1+\delta\,,
\ee
where $\bar{\phi}_1$ is the solution to $f f_1 \dot\phi_2^2 = V_1=0$ and $\delta$ is a small  homogeneous perturbation. One can then study the stability of the solution $\phi_1=\bar{\phi}_1$ by expanding \eqref{eq1} to linear order in $\delta$ 
(we neglect perturbations in $\phi_2$ as we are interested in getting a qualitative picture of the behaviour of the system) 
and solving for the time evolution of the perturbation. From \eqref{eq1} one finds:
\be
\dot\pi_1|_{\bar{\phi}_1}+a^3\left.\left(V_1-f f_1  \dot\phi_2^2\right)\right|_{\bar{\phi}_1}
=-a^3\left(\ddot\delta+3H\dot\delta+\mu^2\delta\right), 
\label{pertur}
\ee
where the mass parameter $\mu$ is defined as:
\be
\mu^2\equiv V_{11}|_{\bar{\phi}_1}-(f_1^2+f f_{11})|_{\bar{\phi}_1} \dot\phi_2^2\,.
\label{eq:mu}
\ee
By definition of $\bar{\phi}_1$, the l.h.s. of \eqref{pertur} vanishes, and so the perturbation to the background trajectory $\delta$ obeys:
\be
\ddot\delta+3H\dot\delta+\mu^2\delta=0\,.
\ee
Evaluating \eqref{eq:mu} on the trivial solution $f f_1 \dot\phi_2^2 = V_1=0$ with $\eta_\perp=0$, we find:
\be
\mu^2= m_{\rm eff}^2|_{\bar{\phi}_1}\,,
\label{eq:mu2}
\ee
indicating that the super-horizon growth of the isocurvature perturbations for $m_{\rm eff}^2<0$ is just an artifact of doing perturbation theory on an unstable background with $\mu^2<0$. This is not a surprise since $\delta$ can be seen as a long wavelength isocurvature perturbation. This means that the trivial solution is not an attractor for the inflationary dynamics, and so the system will reach it only if the initial conditions are finely tuned such that at $t=0$:
\be
\phi_1=\bar{\phi}_1\quad\text{and}\quad \dot{\phi}_1=\ddot{\phi}_1=0\,.
\label{eq:ics}
\ee
However, from a multi-field point of view, the evolution of the system will proceed initially along the steepest directions of the potential without leading to the initial conditions \eqref{eq:ics}. Hence, in general, the system will evolve along the generic solution (\ref{eq:attractorsolu}) which gives $\mu=0$, 
indicating that perturbations get exponentially damped and this non-trivial background is indeed an attractor of the inflationary dynamics.

\begin{figure}[h]
\begin{center}
\includegraphics[width=0.6\textwidth]{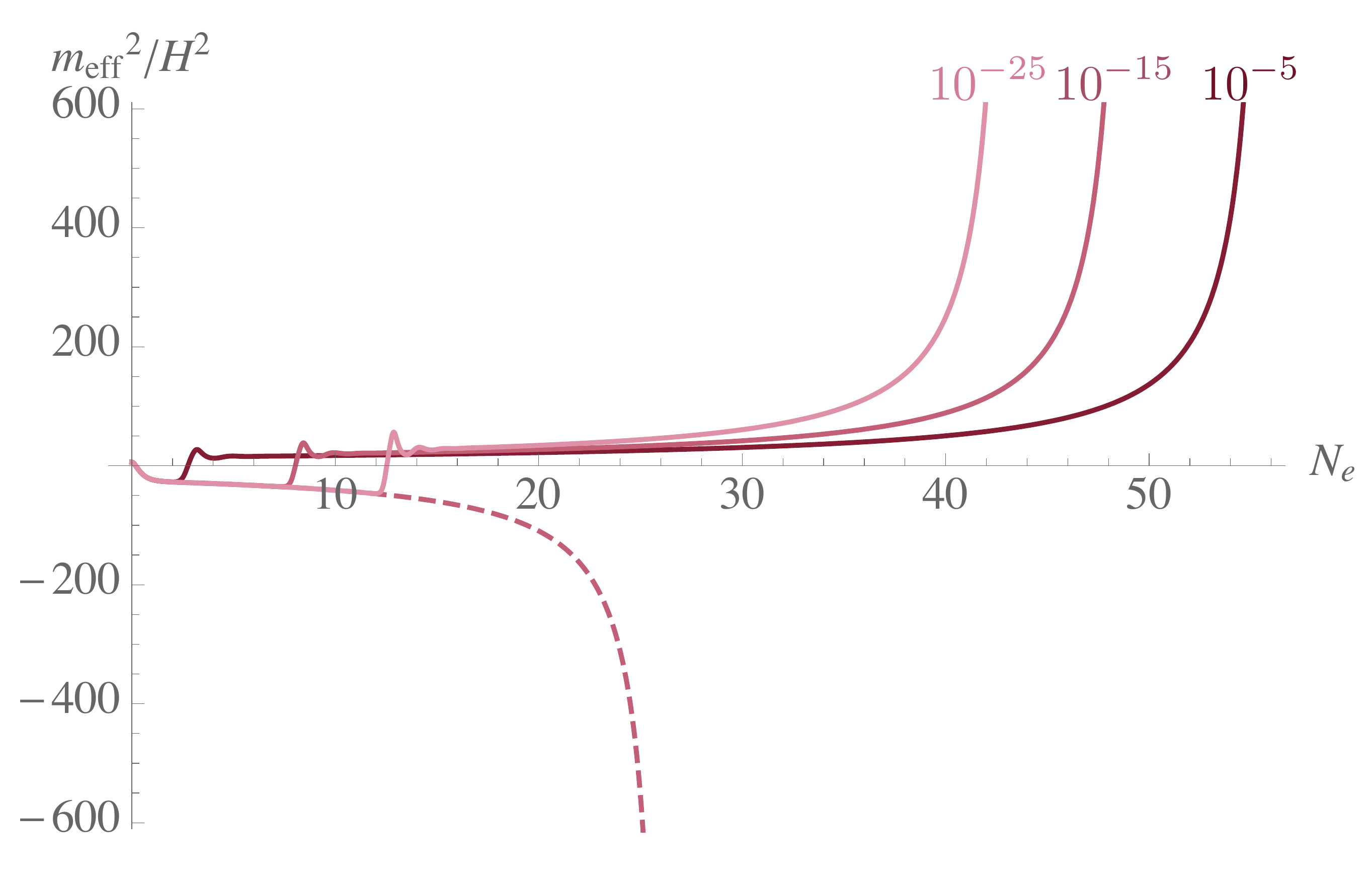}
\caption{Effective mass for the isocurvature modes on the trivial background (dashed) and upon transition from this solution to the attractor (\ref{eq:attractorsolu}) by addition of perturbations to the heavy field $\delta=\{10^{-5},10^{-15},10^{-25}\}$.}
\label{fig:meff2}
\end{center}
\end{figure}

\begin{figure}[h]
\begin{center}
\includegraphics[width=0.60\textwidth]{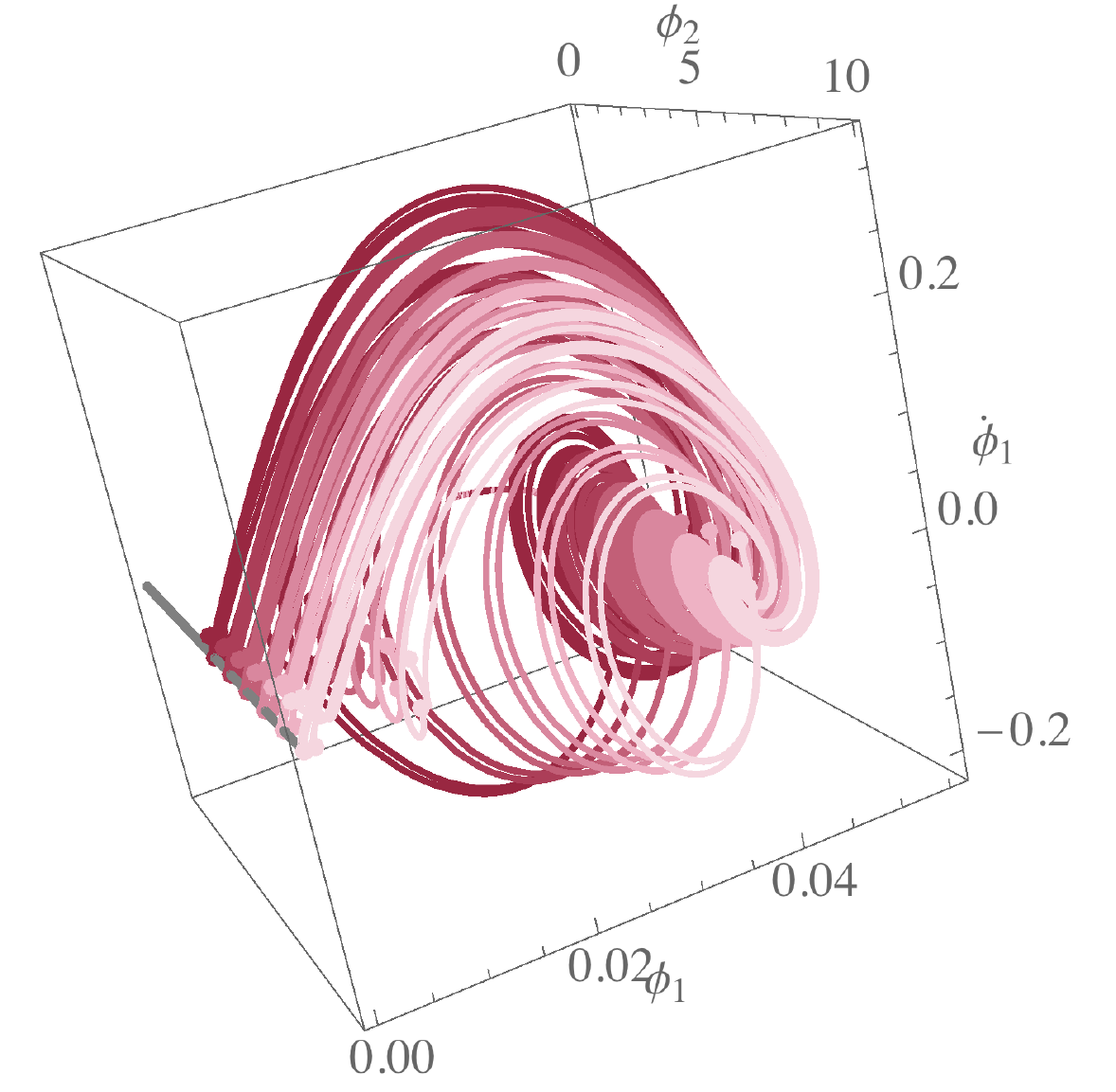}
\caption{Phase space projection of the non-linear sigma model. The trivial trajectory is shown in grey. The repulsive nature of this background trajectory is evident since small perturbations of the initial conditions (\ref{eq:ics}) take the background towards the attractor solution (\ref{eq:attractorsolu}).}
\label{fig:phaseSpace}
\end{center}
\end{figure}

We illustrate this point in Figs. \ref{fig:meff2} and \ref{fig:phaseSpace} which show the dynamics for the minimal geometry of \cite{Renaux-Petel:2015mga}:
\be
f^2(\phi_1)=1+2\frac{\phi_1^2}{M^2}\,,
\ee
supplemented by a double quadratic potential:
\be
V=\frac12 m_1^2 \phi_1^2+\frac12 m_2^2 \phi_2^2\,.
\ee
Given the analytic arguments presented above, the qualitative features of this two-field system do not depend on this particular choice. In fact, similar results can be found with different potentials like that of \cite{Renaux-Petel:2015mga}. In this minimal geometry:
\be
\mathbb{R}=-\frac{4 M^2}{\left(M^2+2 \phi_1^2\right)^2}\simeq -\frac{4}{M^2}\,,
\ee
where in the last step we took $\phi_1/M\ll 1$. By tuning the mass scale $M$ small, one can enhance the effects of the field space curvature and trigger the instability as formulated in \cite{Renaux-Petel:2015mga}. For numerical purposes we have chosen $\{m_1, m_2,M\}=\{1,10,0.05\}$. In Fig. \ref{fig:meff2} we show the evolution of $m_{\rm eff}^2$ for the trivial background where it is always negative, and upon addition of a small perturbation that triggers the transition between the trivial solution $f f_1 \dot\phi_2^2 = V_1=0$ and the attractor of (\ref{eq:attractorsolu}) where $m_{\rm eff}^2>0$. Notice that these results are consistent with what has been previously found in models with non-minimal coupling \cite{Kaiser:2012ak,Schutz:2013fua}.

To highlight the effects of perturbations to the initial conditions (\ref{eq:ics}), we plot in Fig. \ref{fig:phaseSpace} (a subspace of) the phase space of the non-linear sigma model. It is clear that even very small perturbations of (\ref{eq:ics}) take the trajectory away from the trivial solution and into the inflationary attractor, confirming the analytical result of (\ref{eq:mu2}).

\subsection{Canonical inflaton}

We now identify the canonical field $\phi_1$ with the inflaton and study if an instability can arise. The equation for the heavy field $\phi_2$ in (\ref{eq1}) is solved by $V_2\simeq 0$ which from (\ref{eq2}) implies that  $\pi_2 = 0$ is approximately constant with $\dot\phi_2\to 0$ if $f$ does not increase exponentially during inflation. The momentum $\pi_2$ becomes exactly constant for: 
\be
V_2=\dot{\phi}_2=0\,,
\label{eq:sol2}
\ee
implying that on this trivial solution we have $\vec{T}=(\pm 1,0)$ and $\vec{N}=(0,f^{-1}).$
Notice that this solution does not constrain $f_1$, unlike in the case analysed in Sec. \ref{sec:canHeavy}. One can then show that (\ref{eq:sol2}) yields $\eta_\perp=0$ and an effective mass of the form:
\be
m_{\rm eff}^2=\frac{V_{22}}{f^2}+\frac{f_1}{f} V_1-2\frac{f_{11}}{f}\epsilon H^2\,.
\label{eq:meff22}
\ee
Using the slow-roll approximation, the contribution to this effective mass coming from the field space Christoffel symbols can be rewritten as:
\be
\frac{f_1}{f}\,V_1=-3H \frac{f_1}{f} \dot{\phi}_1=-3 H^2 \frac{d \ln f}{d N}\,,
\label{Use}
\ee
where $N=\ln a$ denotes the number of e-foldings. Defining $g(N)\equiv \frac{d \ln f}{d N}$ one can integrate to find:
\be
f(N)=f_0 \,e^{\int_0^N g(N') dN'}\,.
\ee
The isocurvature effective mass (\ref{eq:meff22}) can then be rewritten in terms of the function $g(N)$ as: 
\be
\frac{m_{\rm eff}^2}{H^2}=\frac{V_{22}}{H^2 f^2}-3 g-g^2+g \epsilon -\frac{d g}{dN}\,,
\label{IsoMass}
\ee
which shows that an instability would be present if: 
\be
1\ll\frac{V_{22}}{H^2 f^2} \ll g^2\,.
\label{eq:conds}
\ee
The sign of $g(N)$ is crucial for determining the behaviour of the system. For $g>0$, $f$ grows during inflation and $m_{\rm eff}^2<0$ coincides with the mass of $\phi_2$ becoming sub-Hubble, in contradiction with our assumption that $\phi_2$ is a heavy field which corresponds to the first inequality in (\ref{eq:conds}). In this case one should not talk about an instability since the system becomes effectively a two-field model which would require a different analysis. Conversely, if $g<0$, the contribution to $m_{\rm eff}^2$ coming from the mass of $\phi_2$ increases and prevents the instability from ever taking hold. Thus we conclude that the case with a canonical inflaton does not feature any geometrical instability. 

\section{An instability for ultra-light fields?}
\label{sec:lightscalars}

In Sec. \ref{sec:inst} we have shown that heavy fields with $m_{\rm h}\gg H$ do not suffer from any geometrical destabilisation when the effective mass of the isocurvature perturbations is computed on the attractor background solution. In this section, we shall however point out that the case of \textit{ultra-light} fields with $m_{\rm l}\to 0$ is potentially dangerous since isocurvature fluctuations can become tachyonic when the background trajectory is geodesic. We shall again study separately the two cases where the canonical field is either the ultra-light mode or the inflaton.

\subsection{Canonical ultra-light field}
\label{sec:canLight}

If the ultra-light field is $\phi_1$, $V =V(\phi_2)$ and so $V_1=0$. In this case, as can be seen from (\ref{eq1}), $\pi_1$ is not a conserved quantity and (\ref{eq2}) shows that there is a non-zero turning rate of the background trajectory since $\dot\phi_1\neq 0$. This results in tangent and normal unit vectors with generic non-zero components:
\be
\vec{T}=\frac{1}{|\dot\phi|}\left(\dot\phi_1,\dot\phi_2\right)\qquad \vec{N}=\frac{1}{f |\dot\phi|}\left(-f^2\dot\phi_2,\dot\phi_1\right)\,,
\ee
leading to a non-zero $\eta_\perp$ of the form:
\be
\eta_\perp = \frac{\dot\phi_1 V_2}{H f |\dot\phi|^2}\,.
\ee
Hence the isocurvature mass (\ref{eq:meff2Attractor}) reduces to:
\be
\frac{m_{\rm eff}^2}{H^2 }= 3\eta_\perp^2-2 \epsilon\,\frac{f_{11}}{f}
\simeq 2 \epsilon\left[ \left(\frac{f_1}{f}\right)^2-\frac{f_{11}}{f} \right] +\mathcal{O}(\epsilon^2)\,,
\label{eq:meff2Attractor2}
\ee
where in the last step we have used the slow-roll approximation. Clearly the sign of $m_{\rm eff}^2$ depends on the particular functional dependence of $f(\phi_1)$. A generic supergravity case is $f= f_0\,e^{k \phi_1^p}$ which gives $m_{\rm eff}^2\simeq 0$ for $p=1$ and any value of $k$. This limiting case has been studied in \cite{Cremonini:2010sv, Achucarro:2016fby} which showed that the isocurvature power spectrum remains constant on super-horizon scales and acts as a continuous source of curvature perturbations due to a non-zero coupling induced by $\eta_\perp\neq 0$. Different values of $p$ and $k$ can lead to a positive or negative $m_{\rm eff}^2$, showing that a geometrical instability can potentially arise. Notice that, in contrast with the findings of Sec. \ref{sec:canHeavy}, this case features a genuine instability which is not simply a signal of the repulsive character of the background trajectory since (\ref{eq:meff2Attractor2}) has been computed for $\dot\pi_1 = a^3 f f_1\dot\phi_2^2$ that is the attractor of the dynamical system.

\subsection{Canonical inflaton}

We now study the case where the inflaton is $\phi_1$ while the ultra-light field is $\phi_2$ with $V_2=0$. From (\ref{eq1}) we see that $\pi_2$ is exactly constant, and so in slow-roll (\ref{eq2}) gives:
\be
\dot\phi_2(t) \simeq \dot\phi_2(0) \left(\frac{f(0)}{f(t)}\right)^2\,e^{-3Ht}\,,
\label{dotphi2}
\ee
which shows that $\dot\phi_2(t)=0$ if the initial condition is $\dot\phi_2(0)=0$. In this case the trajectory is exactly geodesic with $\eta_\perp=0$ and the isocurvature mass (\ref{IsoMass}) reduces to:
\be
\frac{m_{\rm eff}^2}{H^2}=-3 g-g^2+g \epsilon -\frac{d g}{dN}\,.
\label{m1}
\ee
This signals the generic appearance of a geometrical instability which could be avoided only for $-3< g<0$. Notice that, using (\ref{Use}), (\ref{m1}) in the slow-roll approximation can also be rewritten as:
\be
\frac{m_{\rm eff}^2}{H^2}=-3\frac{f_1}{f}\frac{\dot{\phi}_1}{H}-2\frac{f_{11}}{f}\epsilon \,,
\ee
showing that the sign of the inflaton velocity is crucial to determine the presence of an instability. We stress again that this would be a genuine instability since we are considering a trajectory which is a dynamical attractor. An interesting string model where such a situation might arise is \textit{Fibre Inflation} \cite{Cicoli:2008gp} where the r\^ole of $\phi_2$ is played by the supersymmetric axionic partner of the inflaton. 

If the initial velocity of the ultra-light field is different from zero but $f$ does not decrease exponentially during inflation, (\ref{dotphi2}) shows that $\dot\phi_2$ relaxes to zero exponentially quickly, and so the previous analysis still holds. Ref. \cite{Krippendorf:2018tei} however considered the case with $V=V_0\,e^{-k_1\phi_1}$ and $f=f_0\,e^{-k_2\phi_1}=f_0\,e^{-k_1 k_2 N}$ where $f$ can decrease exponentially with the number of e-foldings if $k_1 k_2 >0$. In this case $\dot\phi_2$ can no longer be neglected and the system does not evolve along a geodesic trajectory. Hence $m_{\rm eff}^2$ receives a positive contribution proportional to:
\be
\eta_\perp^2 = \left(\frac{f  \dot\phi_2 V_1}{H}\right)^2 \frac{1}{|\dot\phi|^4} \neq 0\,,
\ee
which can prevent the instability. Notice that, when $\dot\phi_2(0)\neq 0$, the system can be studied by integrating out $\dot\phi_2$ and rewriting the first equation in (\ref{eq1}) as:
\be
\dot\pi_1 = - a^3\, V_{{\rm eff},1}\,,
\ee
with a time-dependent effective potential, extending the definition given in \cite{Krippendorf:2018tei} to curved (time varying) backgrounds:
\be
V_{\rm eff} \equiv V(\phi_1) +\frac{\pi_2^2}{2 a^6 f^2(\phi_1)}\,.
\ee

\section{Discussion}

In this paper we have studied the geometrical (in)stability in models of inflation where the field space has negative scalar curvature. These models arise naturally in the presence of non-minimal coupling, in supergravity and in string theory. We have shown that there is no instability for heavy non-inflationary scalars and that the isocurvature modes are tachyonic only in a spurious, non-attractive solution to the background dynamics. Instead we have shown that the instability can be present for massless spectator fields kinetically coupled to the inflaton. When present, this instability should make one reconsider the validity of whatever inflationary model leading to it. The simplest possibility is that perturbation theory remains valid throughout the evolution. In this case the growth of the isocurvature perturbations might lead to a tension with current observational bounds on the isocurvature fraction only if the ultra-light fields contribute considerably to dark matter \cite{Ade:2015lrj}, a possibility which we consider however unlikely given that these fields are in practice massless. One can also envisage a more extreme situation where the growth of the isocurvature perturbations pushes the system beyond the perturbative regime. Should this happen, one would need to resort to more sophisticated approaches where the back-reaction of the perturbations on the background can be incorporated in a consistent fashion, like the stochastic inflation formalism or numerical relativity (for a similar approach during reheating see \cite{Krajewski:2018moi}). We finally mention that the geometrical instability for ultra-light fields with geodesic background trajectory could be avoided by including corrections which give a non-zero mass to the massless modes. However this is not guaranteed to be a successful solution in cases where the geometrical instability can be avoided only if these new mass terms are as large as the inflaton potential since they would severely affect the inflationary dynamics. 

\vspace{0.5cm}
%\paragraph*{\it Acknowledgments:}
\bibliographystyle{JHEP}

\end{document}